# Pseudospin symmetry in the relativistic Killingbeck potential: quasi-exact solution

Majid Hamzavi[1*], Sameer M. Ikhdair[2], Karl-Erik Thylwe[3]

[1]*Department of Basic Sciences, Shahrood Branch, Islamic Azad University, Shahrood, Iran*

[2]*Physics Department, Near East University, 922022, Nicosia, North Cyprus, Mersin 10, Turkey*

[3]*KTH-Mechanics, Royal Institute of Technology, S-100 44 Stockholm, Sweden*

*Corresponding author: Tel.:+98 273 3395270, fax: +98 273 3395270*

Email: majid.hamzavi@gmail.com

## Abstract

The Killingbeck potential consisting of the harmonic oscillator-plus-Cornell potential, $ar^2 + br - c/r$, is of great interest in particle physics. The solution of Dirac equation with the Killingbeck potential is studied in the presence of the pseudospin (p-spin) symmetry within the context of the quasi-exact solutions. Two special cases of the harmonic oscillator and Coulomb potential are also discussed.

**Keywords:** Dirac equation; Killingbeck potential; p-spin symmetry; quasi-exact solution

**PACS:** 03.65.Fd; 02.30.Pm; 03.65.Ge

## 1. Introduction

The particle dynamics in relativistic quantum mechanics is described by using Klein-Gordon (KG) and Dirac equations. The solutions of the relativistic equations are very significant in describing the nuclear shell structure [1-9]. The concept of p-spin symmetry was considered, for the first time, in the non-relativistic about 40 years ago [10-11]. Within the framework of Dirac equation, p-spin symmetry introduced to feature the deformed nuclei and super deformation so that to establish an effective shell-model [12-14]. The spin symmetry is also relevant for mesons [15] and occurs when the difference of the scalar $S(\vec{r})$ and vector $V(\vec{r})$ potentials is a constant, i.e., $\Delta(r) = C_s$ and the p-spin symmetry occurs when the sum of the scalar and vector potentials is a constant, i.e., $\Sigma(r) = C_{ps}$ [16-18].

Since the discovery of Ginocchio, that p-spin symmetry in nuclei can be understood easily in terms of the Dirac equation with large scalar and vector potentials. An increasing number of investigations of the spin- and p-spin symmetry in several physical potential models have been reported [19-29]. We deal with the Dirac equation for the Killingbeck potential to obtain the quasi-exact bound state solutions for relativistic energy eigenvalues and wave functions in the frame of the series expansion method.

The Killingbeck potential is one of these central potential models. It consists of harmonic oscillator plus Cornell potential, i.e., $ar^2 + br - c/r$, which finds applications in particle physics [30,31]. Boumedjane *et al.* investigated the lowest energy states and the corresponding wave functions for the generalized Killingbeck potential within the context of the recently introduced differential quadratic method [32]. Recently, Hamzavi and Rajabi have studied the Dirac equation for the Killingbeck potential under the spin symmetric limit to obtain the energy eigenvalues and the corresponding wave functions by using wave function ansatz method [33].

The purpose of this work is to solve Dirac equation for the Killingbeck potential by quasi-exact solutions under p-spin symmetry. We need to find the bound state solutions including the energy equation and two-component wave function using the quasi-exact method. To this end, we shall first briefly introduce in Section 2 the Dirac equation with radial scalar and vector potentials for arbitrary spin-orbit quantum number $\kappa$ in view of the p-spin symmetry. In Section 3, the general Dirac formulas are then applied to deal with the scalar and vector Killingbeck potentials in order to obtain the energy spectrum equation and the corresponding two-component wave function under the choice of p-spin symmetry. Section 4 is devoted for discussions where we shall consider some particular cases for our solutions like the harmonic oscillator and Coulomb potential. Finally, Section 5 contains summary and concluding remarks.

## 2. Dirac equation under p-spin and spin symmetries

The Dirac equation with scalar $S(r)$ and vector $V(r)$ potentials (in relativistic $\hbar = c = 1$ units) is

$$\left[\vec{\alpha}.\vec{p} + \beta\left(M + S(r)\right)\right]\psi(r) = \left[E - V(r)\right]\psi(r), \tag{1}$$

where $E$ is the relativistic energy of the system, $M$ is the fermionic mass and $\vec{p} = -i\vec{\nabla}$ is the three-dimensional momentum operator [1]. Further, $\vec{\alpha}$ and $\beta$ are the $4 \times 4$ usual Dirac matrices given by

$$\vec{\alpha} = \begin{pmatrix} 0 & \vec{\sigma} \\ \vec{\sigma} & 0 \end{pmatrix}, \quad \beta = \begin{pmatrix} I & 0 \\ 0 & -I \end{pmatrix}, \tag{2}$$

where $\vec{\sigma}$ is Pauli matrix and $I$ is $2 \times 2$ unitary matrix. The total angular momentum operator $\vec{J}$ and spin-orbit $K = (\vec{\sigma}.\vec{L} + 1)$, where $\vec{L}$ is orbital angular momentum, of the spherical nucleons commute with Dirac Hamiltonian. The eigenvalues of spin-orbit coupling operator are $\kappa = \left(j + \frac{1}{2}\right) > 0$ and $\kappa = -\left(j + \frac{1}{2}\right) < 0$ for unaligned spin $j = l - \frac{1}{2}$ and the aligned spin $j = l + \frac{1}{2}$, respectively. $\left(H^2, K, J^2, J_z\right)$ can be taken the complete set of the conservative quantities. Thus, the Dirac spinors can be written according to radial quantum number $n$ and spin-orbit coupling number $\kappa$ as follows

$$\psi_{n\kappa}(\vec{r}) = \begin{pmatrix} f_{n\kappa}(\vec{r}) \\ g_{n\kappa}(\vec{r}) \end{pmatrix} = \frac{1}{r}\begin{pmatrix} F_{n\kappa}(r) Y_{jm}^l(\theta,\phi) \\ iG_{n\kappa}(r) Y_{jm}^{\tilde{l}}(\theta,\phi) \end{pmatrix}, \tag{3}$$

where $f_{n\kappa}(\vec{r})$ is the upper (large) component and $g_{n\kappa}(\vec{r})$ is the lower (small) component of the Dirac spinors. $Y_{jm}^l(\theta,\varphi)$ and $Y_{jm}^{\tilde{l}}(\theta,\varphi)$ are spin and p-spin spherical harmonics, respectively, and $m$ is the projection of the angular momentum on the $z$-axis. Substituting Eq. (3) into Eq. (1) with the usual Dirac matrices, one obtains two coupled differential equations for the upper and the lower radial wave functions $F_{n\kappa}(r)$ and $G_{n\kappa}(r)$ as

$$\left(\frac{d}{dr} + \frac{\kappa}{r}\right) F_{n\kappa}(r) = \left[M + E_{n\kappa} - \Delta(r)\right] G_{n\kappa}(r), \tag{4a}$$

$$\left(\frac{d}{dr} - \frac{\kappa}{r}\right) G_{n\kappa}(r) = \left[M - E_{n\kappa} + \Sigma(r)\right] F_{n\kappa}(r), \tag{4b}$$

where we have used

$$\Delta(r) = V(r) - S(r), \tag{5a}$$

$$\Sigma(r) = V(r) + S(r). \tag{5b}$$

Eliminating $F_{n\kappa}(r)$ and $G_{n\kappa}(r)$ from Eqs. (4), we obtain the following second order Schrödinger-like differential equations for the upper and lower components of the Dirac wave functions, respectively;

$$\left[\frac{d^2}{dr^2} - \frac{\kappa(\kappa-1)}{r^2} - \left(M + E_{n\kappa} - \Delta(r)\right)\left(M - E_{n\kappa} + \Sigma(r)\right)\right]G_{n\kappa}(r)$$
$$-\left[\frac{1}{M - E_{n\kappa} + \Sigma(r)} \frac{d\Sigma(r)}{dr}\left(\frac{d}{dr} - \frac{\kappa}{r}\right)\right]G_{n\kappa}(r) = 0, \tag{6a}$$

$$\left[\frac{d^2}{dr^2} - \frac{\kappa(\kappa+1)}{r^2} - \left(M + E_{n\kappa} - \Delta(r)\right)\left(M - E_{n\kappa} + \Sigma(r)\right)\right]F_{n\kappa}(r)$$
$$+\left[\frac{1}{M + E_{n\kappa} - \Delta(r)} \frac{d\Delta(r)}{dr}\left(\frac{d}{dr} + \frac{\kappa}{r}\right)\right]F_{n\kappa}(r) = 0, \tag{6b}$$

where $\kappa(\kappa-1) = \tilde{l}(\tilde{l}+1)$ and $\kappa(\kappa+1) = l(l+1)$.

## 3. Dirac equation with Killingbeck potential under p-spin symmetry

The exact p-spin symmetry is proved by Meng *et al.* [34-36]. It occurs in the Dirac equation when $\frac{d\Sigma(r)}{dr} = 0$ or $\Sigma(r) = C_{ps} =$ constant. In the presence of this symmetry, from Eq. (6a), one can obtain

$$\left[\frac{d^2}{dr^2} - \frac{\kappa(\kappa-1)}{r^2} - \left(M + E_{n\kappa} - \Delta(r)\right)\left(M - E_{n\kappa} + C_{ps}\right)\right]G_{n\kappa}(r) = 0, \tag{7}$$

where $\kappa = -\tilde{l}$ and $\kappa = \tilde{l}+1$ for $\kappa < 0$ and $\kappa > 0$, respectively. Here, $\Delta(r)$ is to be taken as the Killingbeck potential. Thus, Eq. (7) becomes

$$\left[\frac{d^2}{dr^2} - \frac{\kappa(\kappa-1)}{r^2} - \tilde{\gamma}\left(ar^2 + br - \frac{c}{r}\right) - \tilde{\beta}^2\right]G_{n\kappa}(r) = 0, \tag{8}$$

where

$$\tilde{\gamma} = E_{n\kappa} - M - C_{ps}, \tag{9a}$$

$$\tilde{\beta}^2 = \left(M + E_{n\kappa}\right)\left(M - E_{n\kappa} + C_{ps}\right). \tag{9b}$$

Now, we will solve Eq. (8) for the lower-spinor component $G_{n\kappa}(r)$ and energy spectrum equation. For this, we recast it in the form:

$$\frac{d^2 G_{n\kappa}(r)}{dr^2} + \left[\frac{A_1}{r^2} + \frac{A_2}{r} - A_3 - A_4 r - A_5 r^2\right] G_{n\kappa}(r) = 0, \tag{10}$$

where

$$A_1 = -\kappa(\kappa-1), \; A_2 = \tilde{\gamma}c, \; A_3 = \tilde{\beta}^2, \; A_4 = \tilde{\gamma}b, \; A_5 = \tilde{\gamma}a. \tag{11}$$

The insertion of the ansatz [37,38]

$$G_{n\kappa}(r) = e^{\frac{1}{2}pr^2 + qr} r^\delta \sum_{n=0}^{\infty} a_n r^n, \tag{12}$$

into Eq. (10) leads therefore to the equation

$$\sum_{n=0}^{\infty} a_n X_n r^{n+\delta-2} + \sum_{n=1}^{\infty} a_{n-1} Y_{n-1} r^{n+\delta-2} + \sum_{n=2}^{\infty} a_{n-2} Z_{n-2} r^{n+\delta-2} = 0, \; p^2 = A_5, \; 2pq = A_4, \tag{13a}$$

$$X_n = (n+\delta)(n+\delta-1) + A_1, \tag{13b}$$

$$Y_{n-1} = 2q(n+\delta-1) + A_2, \tag{13c}$$

$$Z_{n-2} = q^2 + 2p(n+\delta-3/2) - A_3, \tag{13d}$$

where $p$ and $q$ are some parameters to be determined by consistency conditions, $G_{n\kappa}(r) \to 0$ when $r \to 0$ and $r \to \infty$. By equating the corresponding power of $r$, we find

$$a_0[\delta(\delta-1) + A_1] = 0, \; a_0 \neq 0,$$

$$a_1 = -\frac{(2q\delta + A_2)}{\delta(\delta+1) + A_1} a_0,$$

$$a_2 = -\frac{\left[(2p\delta + p + q^2) - A_3\right] a_0 + [2q(\delta+1) + A_2] a_1}{(\delta+1)(\delta+2) + A_1},$$

$$a_3 = -\frac{(2pq - A_4) a_0 + [2p(\delta+1) + p + q^2 - A_3] a_1 + [2q(\delta+2) + A_2] a_2}{(\delta+2)(\delta+3) + A_1},$$

.
.
.

$$a_n = -\frac{(p^2 - A_5) a_{n-4} + (2pq - A_4) a_{n-3} + [2p(\delta+n-2) + p + q^2 - A_3] a_{n-2} + [2q(\delta+n-1) + A_2] a_{n-1}}{(\delta+n-1)(\delta+n) + A_1}. \tag{14}$$

The series must be truncated (bounded) for $n = n_r$ that leads to the following equations

$$A_1 = -\delta(\delta-1) = -\kappa(\kappa-1), \; A_2 = -2q(n+\delta-1), \; A_3 = 2p(n+\delta-3/2) + q^2,$$

$$A_4 = 2pq, \; A_5 = p^2. \tag{15}$$

Substituting (11) into (15), we find the parameters in the wave function (12) as

$$\delta = \frac{1}{2} \pm \left(\kappa - \frac{1}{2}\right), \quad p = \pm\sqrt{\tilde{\gamma}}a, \quad q = \pm\frac{\sqrt{\tilde{\gamma}}b}{2\sqrt{a}}, \quad a \neq 0. \tag{16}$$

The negative signs of the coefficients $p = -\sqrt{\tilde{\gamma}}a$ and $q = -\sqrt{\tilde{\gamma}}b/(2\sqrt{a})$ are necessary to ensure square integrability at zero and at ∞, respectively [39] and the positive sign in $\delta = \kappa$ parameter. In addition, by virtue of the relation $A_2 = -2q(n+\delta-1) = c\left(E_{n\kappa} - M - C_{ps}\right)$, one obtains a relationship between the potential parameters:

$$\frac{b^2}{a} = \frac{c^2\left(E_{n\kappa} - M - C_{ps}\right)}{(n+\kappa-1)^2}, \tag{17}$$

and by employing the relation $A_3 = 2p(n+\delta-3/2) + q^2 = (M + E_{n\kappa})(M - E_{n\kappa} + C_{ps})$, together with Eq. (16), one finds

$$(M + E_{n\kappa})(E_{n\kappa} - M - C_{ps}) = 2\sqrt{a}\sqrt{E_{n\kappa} - M - C_{ps}}\left(n + \delta - \frac{3}{2}\right) - \frac{b^2(E_{n\kappa} - M - C_{ps})}{4a}. \tag{18}$$

Therefore, inserting Eq. (17) into Eq. (18), we finally find the energy spectrum formula:

$$(M + E_{n\kappa})(E_{n\kappa} - M - C_{ps}) = 2\sqrt{a}\sqrt{E_{n\kappa} - M - C_{ps}}\left(n + \kappa - \frac{3}{2}\right) - \frac{(E_{n\kappa} - M - C_{ps})^2 c^2}{4(n+\kappa-1)^2}. \tag{19}$$

Some numerical results of the above energy spectrum equation are given in table 1 for different $n$ and $\kappa$. We use parameters as: $a = 0.01, 0.04, 0.1, 0.2\,fm^{-3}$, $c = 1.0$, $M = 5.0\,fm^{-1}$ and $C_{ps} = -5.5\,fm^{-1}$.

The lower component of the wave function from Eq. (12) becomes

$$G_{n\kappa}(r) = Na_0 e^{-\frac{\sqrt{\tilde{\gamma}}}{4\sqrt{a}}(ar^2+br)} r^\kappa$$

$$\times\left\{1 - \frac{(2q\kappa + \tilde{\gamma}c)}{2\kappa}r + \left[\frac{(2q\kappa + \tilde{\gamma}c)(2q(\kappa+1) + \tilde{\gamma}c)}{4\kappa(2\kappa+1)} - \frac{(2p\kappa + p + q^2 - \tilde{\beta}^2)}{2(2\kappa+1)}\right]r^2 - \cdots\right\}, \tag{20}$$

where $p$ and $q$ are given in Eq. (16), $\tilde{\gamma}$ in Eq. (9a) and $N$ is normalization constant. In addition, the upper-spinor component can be found from (4b) as

$$F_{n\kappa}(r) = \frac{1}{(M - E_{n\kappa} + C_{ps})}\left(\frac{d}{dr} - \frac{\kappa}{r}\right)G_{n\kappa}(r).$$

$$= \frac{1}{(M - E_{n\kappa} + C_{ps})} e^{-\frac{\sqrt{\tilde{\gamma}}}{4\sqrt{a}}(ar^2 + br)} r^{\kappa-1} \left( \sum_{n=0}^{\infty} a_n n r^n + \sum_{n=1}^{\infty} a_{n-1} q r^n + \sum_{n=2}^{\infty} a_{n-2} p r^n \right)$$

$$= \frac{1}{(M - E_{n\kappa} + C_{ps})} e^{-\frac{\sqrt{\tilde{\gamma}}}{4\sqrt{a}}(br^2 + ar)} r^{\kappa}$$

$$\times \left[ qa_0 + a_1 + (pa_0 + qa_1 + 2a_2) r + (pa_1 + qa_2 + 3a_3) r^2 \right]. \tag{21}$$

where

$$a_1 = -\frac{(2q\kappa + \tilde{\gamma} c)}{2\kappa} a_0,$$

$$a_2 = \frac{1}{2(2\kappa+1)} \left\{ \frac{1}{2\kappa}(2q\kappa + \tilde{\gamma}c)[2q(\kappa+1) + \tilde{\gamma}c] - \left[(2p\kappa + p + q^2) - \tilde{\beta}^2\right] \right\} a_0,$$

$$a_3 = -\frac{(2pq - \tilde{\gamma}b)}{6(\kappa+1)} a_0 + \frac{(2q\kappa + \tilde{\gamma}c)[2p(\kappa+1) + p + q^2 - \tilde{\beta}^2]}{12\kappa(\kappa+1)} a_0$$

$$- \frac{[2q(\kappa+2) + \tilde{\gamma}c]}{12(\kappa+1)(2\kappa+1)} \left\{ \frac{1}{2\kappa}(2q\kappa + \tilde{\gamma}c)[2q(\kappa+1) + \tilde{\gamma}c] - \left[(2p\kappa + p + q^2) - \tilde{\beta}^2\right] \right\} a_0. \tag{22}$$

## 4. Discussions

In this section, we discuss some special cases. First case, when we set $a = b = 0$, the Killingbeck potential reduces to the Coulomb potential and one can obtain from (19) the energy eigenvalues in the presence of exact p-spin symmetry ($C_{ps} = 0$) as [1]

$$\frac{E_{n,\tilde{l}} + M}{E_{n,\tilde{l}} - M} = -\frac{c^2}{4(n+\tilde{l})^2}, \tag{23}$$

where we have used $\kappa = \tilde{l} + 1$. Second case, when $b = c = 0$, the Killingbeck potential reduces to the harmonic oscillator potential and its energy eigenvalues can be found from (19) as [25,40]

$$(E_{n,\tilde{l}} + M)\sqrt{\frac{E_{n,\tilde{l}} - M}{2M}} = \left(2n_r + \tilde{l} + \frac{3}{2}\right)\omega, \tag{24}$$

where we used $a = \frac{1}{2} M \omega^2$, $\kappa = \tilde{l} + 1$ and a new quantum number $n = 2(n_r + 1)$.

## 5. Summary and Concluding Remarks

In this work, we solved the Dirac equation with the central Killingbeck potential in view of the p-spin symmetric limit for any arbitrary spin-orbit coupling quantum number $\kappa$ using the quasi-exact method. Also, the numerical results for the p-spin energy levels are calculated using a set of parameter values in Table 1. The bound state energy spectrum equation and the two-spinor component radial wave function are obtained. The radial wave function is square integrable at zero and at $\infty$, respectively. We also discussed two simple cases including the harmonic oscillator and the Coulomb potential. They are in agreement with those found before by other authors.

## Acknowledgment

The authors thank the referees for their invaluable suggestions that have helped in improving this paper.

**Table 1:** The bound state energy eigenvalues under p-spin symmetry case for Killingbeck potential when $c = 1$ [32].

| $n, \kappa$ | $a$ | $b$ | $E_{n,\kappa}$ ($fm^{-1}$) |
|---|---|---|---|
| 1, -1 | 0.01 | 0.0000443352 | -0.4955664823 |
|  | 0.04 | 0.0007042067 | -0.4823948332 |
|  | 0.10 | 0.0043392099 | -0.4566079010 |
|  | 0.20 | 0.0042704702 | -0.4145905955 |
| 1, -2 | 0.01 | 0.0000306861 | -0.4877255699 |
|  | 0.04 | 0.0004827590 | -0.4517240996 |
|  | 0.10 | 0.0029228264 | -0.3830869446 |
|  | 0.20 | 0.0049669143 | -0.2764888562 |
| 2, -1 | 0.01 | 0.0000136408 | -0.4877232641 |
|  | 0.04 | 0.0002147143 | -0.4516892799 |
|  | 0.10 | 0.0013012046 | -0.3828915833 |
|  | 0.20 | 0.0049669143 | -0.2764888562 |
| 2, -2 | 0.01 | 0.0000149613 | -0.4760618300 |
|  | 0.04 | 0.0002321431 | -0.4071427474 |
|  | 0.10 | 0.0013728526 | -0.2803435787 |
|  | 0.20 | 0.0050762262 | -0.0939018983 |